# Ultrashort waveguide tapers based on Luneburg lens

S. Hadi Badri[a,*], M. M. Gilarlue[b]

[a] Young Researchers and Elite Club, Tabriz Branch, Islamic Azad University, Tabriz, Iran

[b] Department of Electrical Engineering, Sarab Branch, Islamic Azad University, Sarab, Iran

* sh.badri@iaut.ac.ir

## Abstract

In integrated photonic circuits, silicon-on-insulator waveguides with different geometries have been employed to realize a variety of components. Therefore, efficient coupling of two different waveguides is crucial. In this paper, focusing property of the Luneburg lens is exploited to design waveguide tapers. The Luneburg lens, truncated in a shape of a parabolic taper with reduced footprint, is utilized to connect a 10 μm-wide waveguide to a 0.5 μm one with the same thickness with an average coupling loss of 0.35 dB in the entire O, E, S, C, L, and U bands of optical communications. The proposed compact taper with the length of 11 μm is implemented by varying the thickness of the guiding layer and compared with three conventional tapers with the same length. However, designing a coupler to connect waveguides with different thicknesses and widths is more challenging. By applying quasi-conformal transformation optics, we flatten the Luneburg lens and consequently increase the refractive index on the flattened side. As a result, we are able to couple two waveguides with different thicknesses and widths. The numerical simulations are used to evaluate the theoretically designed tapers. To our knowledge, this is the first study presenting ultrashort tapers based on truncated Luneburg lens.

## 1. Introduction

In photonic integrated circuits (PIC), the silicon-on-insulator (SOI) waveguides in different components may have different widths and/or thicknesses to realize diverse functionalities such as sensors [1, 2], Mach-Zehnder interferometers [3, 4], modulators [5, 6], ring isolators [7, 8], and grating couplers [9, 10]. Consequently, efficient coupling between different optical waveguides is required in many devices. Connecting components with different waveguide geometries requires a low-loss taper with a compact footprint. Moreover, the fundamental mode should propagate through tapers without converting into higher-order modes [11]. Tapers, also called spot-size converters, have been employed in a variety of devices such as arrayed-waveguide gratings [12], multimode interference couplers [13], and waveguide crossings [14]. Ensuring the adiabatic operation of the taper requires that reducing the width of the taper be slower than the diffraction spreading of the first-order mode, therefore, adiabatic tapers are considerably long. An adiabatic taper with a length of 120 µm has been reported to connect a 12 µm-wide waveguide to a 0.5 µm one with a coupling loss lower than 0.07 dB [15]. An adiabatic 22.5 µm-long taper based on a flat lens has been proposed to connect a 10 µm-wide waveguide to a 0.5 µm one with coupling loss lower than 0.45 dB in a bandwidth of 60 nm [16]. A taper using a quadratic sinusoidal function has been designed to connect a 10 µm-wide waveguide to a 1 µm one with a length of 19.5 µm in silicon nitride (SiN) platform. This study reports a maximum 0.27 dB coupling loss in the C and L bands [17]. The same method has also been utilized in SOI platform to design a 15 µm long taper with a coupling loss lower than 0.31 dB for connecting a 10 µm-wide waveguide to a 0.5 µm single-mode silicon waveguide [18]. Segmented-stepwise taper with a length of 20 µm has been designed to connect a 12 µm-wide waveguide to a 0.5 µm one with a maximum coupling loss of 0.5 dB [19]. A hollow tapered waveguide has been introduced to reduce the lateral width of the waveguide from 15 µm to 0.3 µm at a length of 60 µm while the coupling loss is lower than 2.0 [20]. The theoretical results of [15-20] are reported in this section. Tapers based on transformation optics (TO) have also been proposed, however, the implementation of these designs require anisotropic metamaterials [21-23].

Recently, gradient index (GRIN) lenses such as Eaton [24, 25], Luneburg [26, 27], and Maxwell's fisheye [28, 29] lenses have found wide application in PICs. Luneburg lens focuses the parallel rays incident on its side to a point on the opposite side. In this paper, the focusing property of the Luneburg lens is utilized to design compact high-efficiency waveguide tapers in SOI platform. To the best of our knowledge, this is the first study that investigates the design of ultrashort tapers based on truncated Luneburg lens, facilitating the realization of dense PICs. The lenses are truncated in a shape of a parabolic taper to reduce the footprint of the designed tapers. The performance of the proposed tapers, which connect SOI waveguides with different geometries, are evaluated by numerical simulations. In section 2, we truncate the circular Luneburg lens to reduce its footprint and couple a 10 µm-wide waveguide to a 0.5 µm one with the same thickness. The coupling loss of this taper with a length of 11 µm is lower than 0.87 dB in the 1260-1675 nm bandwidth. In section 3, we flatten one side of the Luneburg lens with transformation optics (TO) to increase the refractive index at its side. We present four designs in this section to show that a variety of waveguides with different widths and thicknesses can be coupled efficiently with this method. For instance, a waveguide of 10 µm × 0.11 µm (width×thickness) is coupled to a 0.5 µm × 0.29 µm waveguide. The maximum coupling loss of 1.02 dB is achieved for this 10.95 µm long taper in the 1260-1675 nm bandwidth. In section 4, possible implementations for the designed tapers are presented based on varying the thickness of the guiding layer. Furthermore, the designed tapers are compared with previous studies.

## 2. Circular Luneburg lens as waveguide taper

The refractive index profile of the generalized Luneburg lens is described by [30]

$$n_{lens}(r) = n_{edge}\sqrt{1+f^2-(r/R_{lens})^2}/f \quad , \quad (0 \leq r \leq R_{lens}) \tag{1}$$

where $n_{edge}$ is the refractive index of the lens at its edge, $r$ is the radial distance from the center, $R_{lens}$ is the radius of the lens, and $f$ determines the position of the focal point. For $f=1$, the focal point lies on the edge of the lens while for $f>1$ and $f<1$ the focal point of the lens is located outside and inside of the lens, respectively. In this study, we use $f=1$ in our calculations of the refractive indices of the studied Luneburg lenses.

In this section, the circular Luneburg lens is used to couple SOI waveguides with the same thickness but with different widths. The SOI waveguides with silicon (Si) guiding layer, silica substrate, and air cladding are considered in this study. When the thickness of Si is 110 nm, the effective refractive index of the slab waveguide is about 2.2. In this section, the Si layer in all the waveguides is $h=110$ nm thick. The effective refractive indices of the waveguides, as well as the Luneburg lens with $L_a = 2R_{lens} = 11\mu m$ and $n_{edge} = 2.2$, are shown in Fig. 1(a). The width of waveguides are $w_{in} = 10\mu m$ and $w_{out} = 0.5\mu m$. The electric field distribution of the TE$_0$ mode at the wavelength of 1550 nm for the circular Luneburg taper is shown in Fig. 1(b) while the coupling loss of 0.53 dB and return loss of 15.2 dB are achieved at this wavelength. We translate the refractive indices given in the conceptual design of Fig. 1(a) to the thickness of the silicon slab waveguide. We implement the lens by mapping the designed two-dimensional (2D) refractive index distribution to the three-dimensional (3D) thickness profile of silicon layer. In order to achieve impedance matching at the interface of the waveguide and the lens, we should avoid introducing any sharp changes in the thickness or width of the silicon layer to minimize reflection at the interface. Therefore, the thicknesses of the waveguides are also chosen based on the same mapping to avoid step-like changes in the thickness of the designed tapers, i.e., the effective refractive indices of the slab waveguides are used in the designs presented in this paper and the effect of the waveguide's width on its effective refractive index is not considered. Moreover, the lens is truncated as shown in Fig. 1(c), so the width of the silicon layer changes gradually throughout the designed taper. Consequently, the thicknesses and widths of the waveguides and the lens's edge are the same, therefore, the impedance matching at the interface of the lens and the waveguides is achieved. The data provided in [31] are used to map the effective refractive index to guiding layer's thickness. The implementation method is further discussed in section 4. We truncate the Luneburg lens based on the following interpolation formula [17, 18]:

$$x = a(by^2 + (1-b)y) + (1-a)\sin\frac{c\pi y^2}{2} \tag{2}$$

where $0 \leq a \leq 1$, $-\frac{c}{c-2} \leq b \leq \frac{c}{c-2}$, $c$ is an odd integer $\geq 3$, and $y$ is the relative length of the taper. $c$ controls the number of full oscillations of the sinusoidal component part of the taper while $a$ controls the fraction of the sinusoidal and parabolic component. And b controls the parabolic curvature of the baseline [18]. In the truncated lens, $a=b=1$ and $c=3$ is used. Fig 1(c) displays the electric field distribution in the truncated Luneburg lens as a taper. Since the lens is truncated in the shape of the parabolic taper, it benefits from its tapering effect and its performance is slightly better than the complete lens. In this case, the return loss is 16.6 dB while the coupling loss is 0.35 dB at the wavelength of 1550 nm. We also truncated the Luneburg lens in the shape of linear and Gaussian tapers, however, the lens truncated in the shape of the parabolic taper had the best performance. Hereafter, we refer to the truncated Luneburg taper of Fig. 1(c) simply as Taper A.

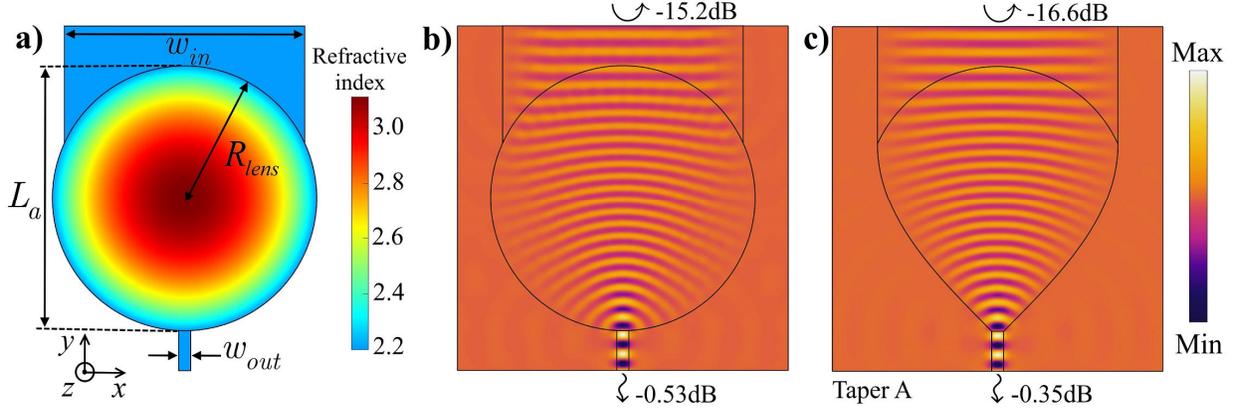

Fig. 1. a) The Luneburg lens couples a waveguide with wider width ($w_{in} = 10\mu m$) to a narrower one ($w_{out} = 0.5\mu m$). The length of the taper is $L_a = 11\mu m$ and thickness of waveguides is $h$=110 nm. b) The electric field distribution of the $TE_0$ mode in the complete Luneburg lens. c) The lens is truncated to reduce the footprint of the taper.

The performance of Taper A is compared with three types of tapers with the same length ($L_a = 11\mu m$). Fig. 2 shows the electric field distribution of the parabolic, linear, and Gaussian tapers at a wavelength of 1550 nm. The parabolic taper, shown in Fig. 2 (a), is obtained $a=b=1$ and $c=3$. For linear taper, $a=1$, $b=0$ is used, while Gaussian taper corresponds to $a=0$, $c=1$. The waist of the field wavefronts is effectively reduced along Taper A, while conventional tapers of Fig. 2 are not able to narrow the wide waist of the field wavefronts in such short lengths. Consequently, the wavefront of the optical wave is considerably distorted in the tapers of Fig. 2 resulting in mode distortion and considerable coupling loss. The coupling losses of the parabolic, linear, and Gaussian tapers are 6.8, 4.7, and 5.1 dB while the return losses are 8.7, 20, 10.1 dB, respectively.

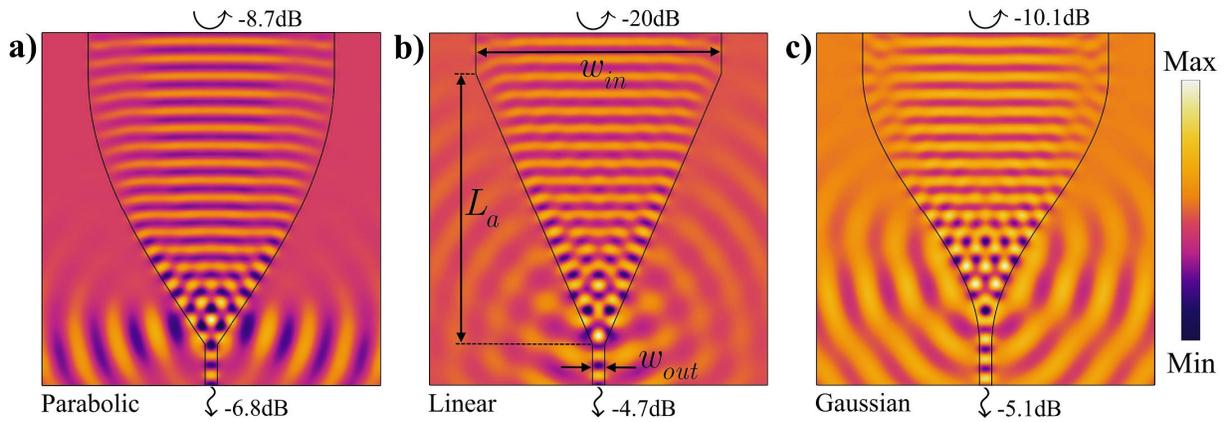

Fig. 2. The electric field distribution of the $TE_0$ mode in the three types of tapers formed by Eq. (2). a) Parabolic taper where $a=b=1$ and $c=3$. b) Linear taper where $a=1$, $b=0$. c) Gaussian taper where $a=0$, $c=1$. The width of input

and output waveguides are $w_{in} = 10 \mu m$ and $w_{out} = 0.5 \mu m$, respectively. The length of the tapers is the same as Taper A ($L_a = 11 \mu m$) while $h$=110 nm.

We also compare the performance of Taper A with the linear taper with the same length and design for the second-order mode. However, in order to support the TE$_1$ mode, the widths of the output waveguides are increased to 1 μm. We refer to this taper as Taper A′. The electric field distributions of the TE$_1$ mode for Taper A′ and linear taper are illustrated in Fig. 3. For higher order modes, the performance of the conventional tapers degrades considerably compared to Taper A′ with the same length. Taper A′ has a coupling loss of 0.34 dB while the linear taper has a coupling loss of 7.8 dB for the TE$_1$ mode at a wavelength of 1550 nm. The return losses are 16.5 and 20.4 dB for Taper A′ and the linear taper, respectively.

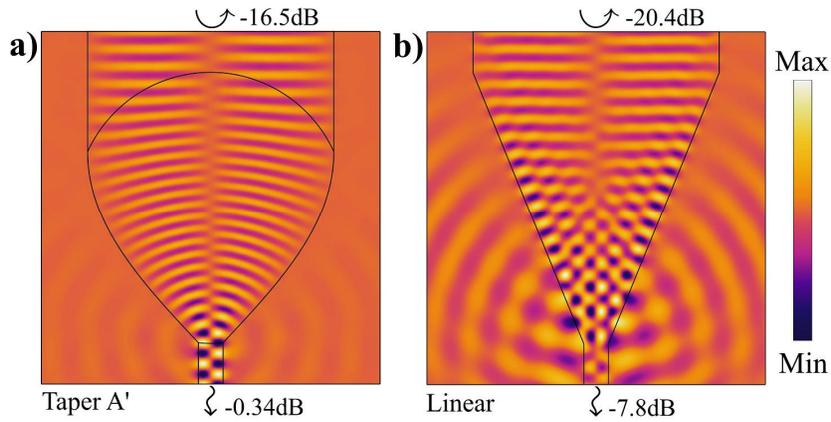

Fig. 3. The electric field distribution of the TE$_1$ mode for a) Taper A′ and b) linear taper. The width of input and output waveguides are $w_{in} = 10 \mu m$ and $w_{out} = 1 \mu m$ while the length of the coupler is $L_a = 11 \mu m$.

## 3. Flattened Luneburg lens as waveguide taper

The circular Luneburg lens can only couple waveguides with the same effective refractive index to each other, i.e., with the same thickness. However, there may be cases where the thicknesses of the waveguides are different. For instance, when the thickness of a waveguide is 110 nm while the other one is 250 nm, the circular lens cannot be used to couple them. In [1-10], the thickness of the SOI waveguides range from 40 to 380 nm, therefore, we present designs to couple waveguides with different thicknesses in this section.

TO provides a method to transform the known geometry (virtual domain) to an arbitrary geometry (physical domain) with the same optical response. Tapers designed by TO may require anisotropic metamaterials for their implementation [21-23]. Therefore, we apply quasi-conformal TO (QCTO) to flatten the Luneburg lens, making it possible to implement it by conventional isotropic materials. Flattening of the Luneburg lens based on QCTO has been studied [32, 33]. Hence, we briefly describe the QCTO method in this paper. The circular Luneburg lens, virtual domain, is shown in Fig. 4(a). The orthogonal grid is created by solving the Laplace equation for two times. Initially, we apply Dirichlet and Neumann boundary conditions to the blue and red boundaries, respectively. Then we solve the Laplace equation for the second time by applying Dirichlet and Neumann boundary conditions to the red and blue boundaries, respectively. Consequently, the generated grids are orthogonal to the boundaries. Two quadrilaterals can be mapped conformally onto each other provided that they have the same conformal module, M. Conformal module is the ratio of the lengths of the two adjacent sides of the quadrilateral. We map the circular virtual domain onto an intermediate domain which is a rectangle with the same M. Then the intermediate domain is mapped onto the physical domains of Figs. 4(b), 4(c), and 4(d). The intermediate domain is not shown in Fig. 4. It is possible to design a large number of physical domains by modifying M. In the virtual domain, M can be adjusted by changing $\theta$; where $\theta$ determines the length of Dirichlet and Neumann boundaries. The point in the lower corner of the physical domain $(x_p, y_p)$ determines its shape and consequently its conformal module. The refractive index profile of the physical domain can also be modified by changing the refractive index profile of the virtual domain, i.e., $n_{edge}$. The details of design parameters for each of the physical domains and the tapers designed based on them are discussed in following subsections.

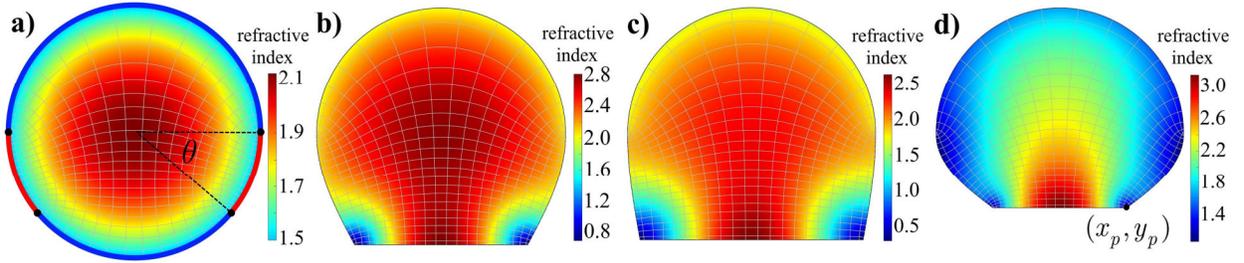

Fig. 4. a) The virtual domain is a circular Luneburg lens where the Dirichlet–Neumann boundaries are shown in blue and red lines. It is possible to couple waveguides with a large range of effective refractive index by changing $n_{edge}$, the geometry of the physical domain, and the lengths of the Dirichlet–Neumann boundaries. The physical domains of b), c), and d) are used to design Tapers B, C, and D, respectively.

## 3.1 Taper B

We refer to the taper designed based on the lens of Fig. 4(b) as Taper B. The design parameters in the virtual domain are $R_{lens} = 2.5 \mu m$, $n_{edge} = 2.0$, and $\theta = 50°$. In the physical domain of Fig. 4(b), the design parameters are $x_p = 0.695 \times R_{lens}$ and $y_p = -0.889 \times R_{lens}$. As shown in Fig. 5(a), the lens of Fig. 4(b) is truncated as a parabolic taper with Eq. 2 where the connecting waveguides with refractive indices (thicknesses) of 2.0 (90 nm) and 2.8 (250 nm) are also displayed. The widths of waveguides are $w_{in} = 4 \mu m$ and $w_{out} = 0.8 \mu m$ while the length of the taper is $L_b = 4.72 \mu m$. The electric field distribution of the TE$_0$ mode in Taper B is displayed at 1550 nm in Fig. 5(b) while the return and coupling losses are 28 dB and 0.07 dB, respectively, at this wavelength. In the C-band, the coupling loss is lower than 0.074 dB while the coupling loss is lower than 0.13 dB in the entire optical communication bands (i.e., O, E, S, C, L, and U bands) for Taper B.

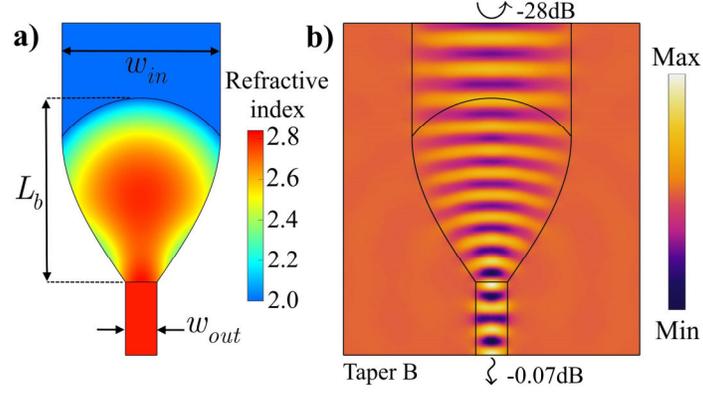

Fig. 5. a) The refractive indices of the coupled waveguides are 2.0 and 2.8 while $w_{in} = 4\mu m$, $w_{out} = 0.8\mu m$, and $L_b = 4.72\mu m$ b) The electric field distribution of TE$_0$ mode in Taper B at a wavelength of 1550 nm.

### 3.2 Taper C

The taper designed based on the lens of Fig. 4(c) is named as Taper C. The design parameters in the virtual domain are $R_{lens} = 2.5\mu m$, $n_{edge} = 1.7$, and $\theta = 40°$ while in the physical domain the design parameters are $x_p = 0.900 \times R_{lens}$ and $y_p = -0.853 \times R_{lens}$. The truncated lens of Fig. 4(c), truncated in a shape of a parabolic taper with Eq. 2, is shown in Fig. 6(a). The waveguides with effective refractive indices (thicknesses) of 1.7 (60 nm) and 2.6 (170 nm) are also displayed in this figure. The widths of waveguides are $w_{in} = 4\mu m$ and $w_{out} = 0.8\mu m$ while the length of the taper is $L_c = 4.63\mu m$. The electric field distribution of the TE$_0$ mode in Taper C is shown in Fig. 6(b) while the return and coupling losses are 27 dB and 0.29 dB, respectively, at the wavelength of 1550 nm. The coupling loss is lower than 0.31 dB in the C-band. However, the coupling loss of less than 0.55 dB is achieved in the entire optical communication.

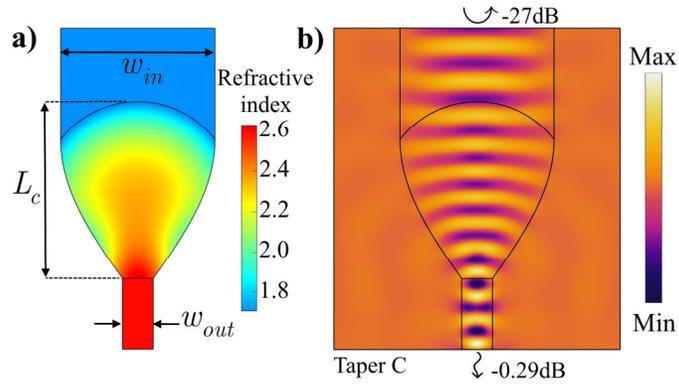

Fig. 6. a) The refractive indices of the coupled waveguides are 1.7 and 2.6 while $w_{in} = 4\mu m$, $w_{out} = 0.8\mu m$, and $L_c = 4.63\mu m$ b) The electric field distribution of TE$_0$ mode in Taper C at a wavelength of 1550 nm.

### 3.3 Taper D

The taper designed based on the lens of Fig. 4(d) is named as Taper D. The design parameters in the virtual domain are $R_{lens} = 2.5\,\mu m$, $n_{edge} = 1.65$, and $\theta = 40°$. In the physical domain of Fig. 4(d), the design parameters are $x_p = 0.542 \times R_{lens}$ and $y_p = -0.600 \times R_{lens}$. The truncated lens of Fig. 4(d), as well as the connecting waveguides with effective indices (thicknesses) of 1.65 (55 nm) and 3.1 (350nm), are also displayed in Fig. 7(a). The lens is truncated in a shape of a parabolic taper with Eq. 2. The widths of waveguides are $w_{in} = 4\,\mu m$ and $w_{out} = 0.8\,\mu m$ while the length of the taper is $L_d = 4.00\,\mu m$. The electric field distribution of the TE$_0$ mode in Taper D is displayed in Fig. 7(b) while the return and coupling losses are 26 and 0.39 dB, respectively, at the wavelength of 1550 nm. In the C-band, the coupling loss lower than 0.41 dB is achieved while the coupling loss is lower than 0.69 dB in the entire optical communication bands; i.e., O, E, S, C, L, and U bands.

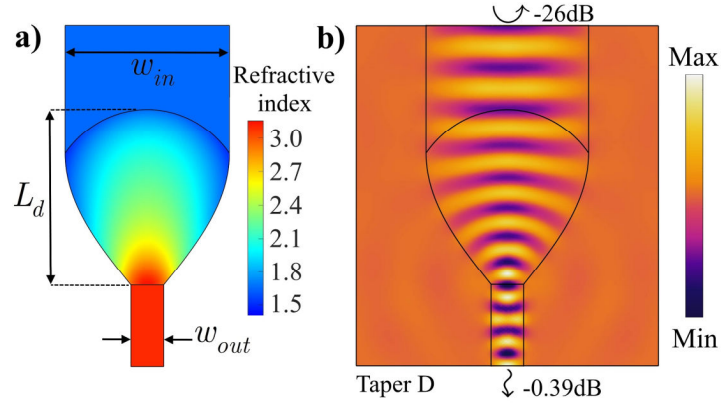

Fig. 7. a) The refractive indices of the coupled waveguides are 1.65 and 3.1 while $w_{in} = 4\,\mu m$, $w_{out} = 0.8\,\mu m$, and $L_d = 4.00\,\mu m$ b) The electric field distribution of TE$_0$ mode in Taper D at a wavelength of 1550 nm.

### 3.4 Taper E

The last taper is named as Taper E. The physical domain for this design is not shown in Fig. 4. The design parameters in the virtual domain are $R_{lens} = 5.75\,\mu m$, $n_{edge} = 2.2$, and $\theta = 55°$. In the physical domain, the design parameters are $x_p = 0.600 \times R_{lens}$ and $y_p = -0.905 \times R_{lens}$. The truncated lens, truncated in a shape of a parabolic taper with Eq. 2, is shown in Fig. 8(a). In this figure, the connecting waveguides with effective indices (thicknesses) of 2.2 (110 nm) and 3.0 (290 nm) are also displayed. The widths of waveguides are $w_{in} = 10\,\mu m$ and $w_{out} = 0.5\,\mu m$ while the length of the taper is $L_e = 10.95\,\mu m$. The truncated transformed Luneburg lens effectively compresses the electromagnetic field along the taper allowing efficient coupling of the two waveguides [Fig. 8(b)]. For the TE$_0$ mode, the return and coupling losses are 23 and 0.16 dB, respectively, at the wavelength of 1550 nm. The effective refractive indices of 2.2 and 3.0 correspond to waveguides with Si thickness of about 110 nm and 290 nm, respectively. The performance of Taper E is discussed in the following section.

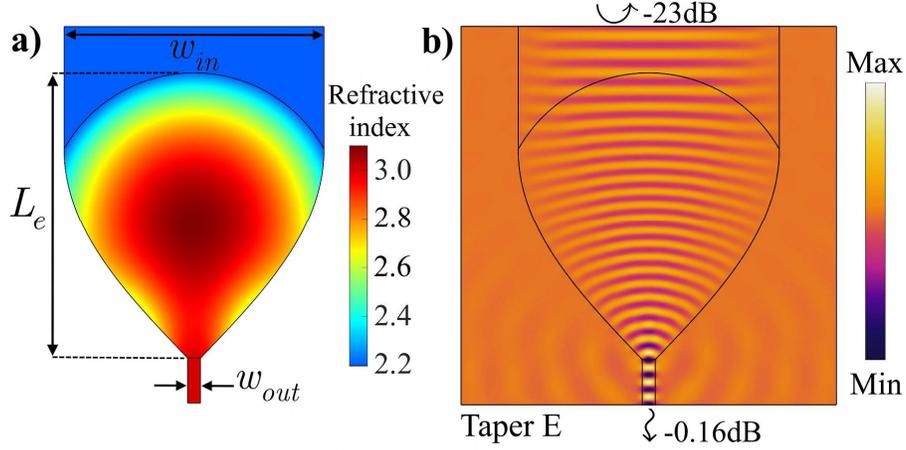

Fig. 8. a) The refractive indices of the coupled waveguides are 2.0 and 3.0 while $w_{in} = 10 \mu m$, $w_{out} = 0.5 \mu m$, and $L_e = 10.95 \mu m$ b) The electric field distribution of $TE_0$ mode in Taper E at a wavelength of 1550 nm.

## 4. Results and discussion

The 2D simulations are performed by Comsol Multiphysics ® for QCTO, generating electric field distribution figures, and creating 3D models of the designed lenses. In 2D simulations, the ideally designed 2D refractive indices, shown in Figs. 1(a), 5(a), 6(a), 7(a), 8(a), surrounded by air are used. Since the 3D finite difference time domain (FDTD) simulations require less memory, Lumerical FDTD Solutions ® software is used to calculate the scattering parameters of the designed tapers. The 3D models of the lenses created by Comsol (exported as STL files) are imported to Lumerical for 3D simulations. The built-in Si and $SiO_2$ material models of Lumerical software are used in these calculations. The maximum meshing step in lateral direction (*x* and *y* axes) is 25 nm while in the vertical direction (*z* axis) is 10 nm. The *x*, *y*, and *z* axes are only shown in Fig. 1(a). A mode source is used to inject a TE mode signal into the simulation region. GRIN lenses have been implemented by graded photonic crystals [34], subwavelength gratings [35, 36], multilayer structures [37, 38], and varying the thickness of guiding layer in a slab waveguide [31, 39]. Gray-scale E-beam lithography has been used to fabricate GRIN lenses by varying the silicon layer's thickness [26, 27, 40, 41]. We implemented all the designed tapers based on varying the thickness of the guiding layer and evaluated their performance numerically, however, we only present the implementations of Tapers A and E in this paper. The reported return and coupling losses in Figs. 1-3 and 5-8 are based on 3D simulations. The procedure to implement a designed refractive index profile based on varying the thickness of the guiding layer in a slab waveguide is described in [31, 40, 41]. In order to convert the 2D gradient index distribution of the conceptual designs of Figs. 1(a), 5(a), 6(a), 7(a), and 8(a) to the 3D thickness profile of the silicon layer, we consider a silicon slab waveguide with a fixed width where the upper and lower claddings are air and silica, respectively. Therefore, the effective refractive index of the slab waveguide only depends on the thickness of the silicon guiding layer. We calculate the effective refractive index of the slab waveguide for different thickness values and fit them to a curve similar to the ones provided in [31, 41]. Finally, we translate the 2D gradient index distribution of the designed lenses to the 3D thickness profile of the silicon layer based on the fitted curve. The effect of the width in translation from the 2D design to the 3D thickness profile is not considered similar to previously fabricated GRIN lenses [26, 40, 41]. The thicknesses of the coupled waveguides are determined with the same fitted curve, consequently, the thicknesses of the waveguides and the lens at their interfaces match. If we consider the width of the waveguides in effective index calculations, the required silicon thickness to satisfy the designed effective refractive indices of the waveguides will be different from the thickness of the lens at its edges resulting in considerable reflection from these interfaces. Another factor that introduces reflection is sharp changes in the width of the silicon guiding layer. We truncate the designed lenses, therefore, the width of the silicon layer changes gradually throughout the designed tapers. The truncation of the lens also reduces the footprint of the designed tapers. Our design concept combined with metasurfaces may be used to manipulate the wavefront [42, 43].

### 4.1 Implementation of Taper A

The implementation of Taper A of Fig. 1(c) is shown in Fig. 9. In this figure, the Si guiding layer and SiO$_2$ cladding are shown while the upper air cladding is not displayed. The performance of the implemented Taper A is compared with the conventional tapers of Fig. 2 based on 3D simulations. The width of the input waveguide is $w_{in} = 10\mu m$ while for the TE$_0$ and TE$_1$ modes the widths of the output waveguides are $w_{out} = 0.5\mu m$ and $w_{out} = 1\mu m$, respectively. The thickness of the guiding layer is $h$=110 nm in the waveguides. As shown in Fig. 10, the coupling loss is lower than 0.38 dB in the C-band. The coupling losses of conventional tapers are considerably higher. Linear taper has lower loss compared with parabolic and Gaussian tapers. The coupling loss of the linear taper is lower than 4.75 dB in the C-band. When the difference between the widths of the waveguides is large, conventional tapers should be considerably longer to achieve adiabatic propagation.

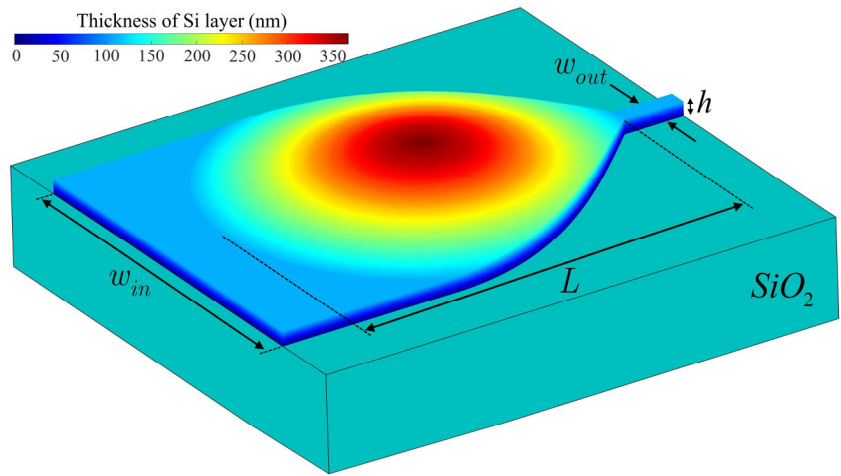

Fig. 9. The implementation of Taper A [Fig. 1(c)] with $L = 11\mu m$, $w_{in} = 10\mu m$, $w_{out} = 0.5\mu m$ and $h$=110 nm. The upper air cladding is not displayed.

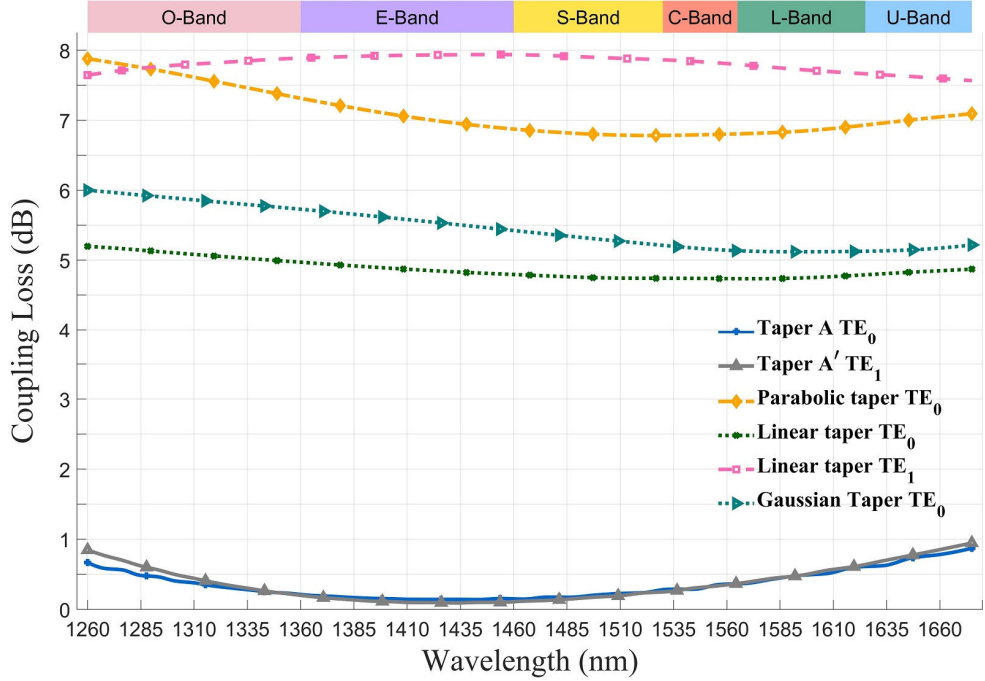

Fig. 10. The performance of Taper A is compared with conventional tapers of equal length based on 3D simulations. For the TE$_1$ mode, tapers of Fig. 3 with $w_{out} = 1\mu m$ are considered.

### 4.2 Implementation of Taper E

As shown in Fig. 11(a), Taper E is implemented by varying the thickness of the guiding layer. This taper connects a waveguide with $w_{in} = 10\mu m$ and $h_{in} = 110nm$ to the output waveguide with $w_{out} = 0.5\mu m$ and $h_{out} = 290nm$. We also display a linear taper to connect the same waveguides to each other in Fig. 11(b). Both tapers have a length of $L_e = 10.95\mu m$. The coupling loss of these tapers are shown in Fig. 12. For Taper E, the average coupling loss is 0.30 dB in the entire O, E, S, C, L, and U bands of optical communications. In the C-band, the coupling loss of less than 0.37 dB is achieved. The coupling losses of the linear taper of Fig. 11(b) is higher than 4.45 dB. We also connected the above waveguides with parabolic and Gaussian tapers. In the bandwidth of 1260-1675 nm, the coupling losses are higher than 7.3 dB and 5.6 dB for the parabolic and Gaussian tapers, respectively.

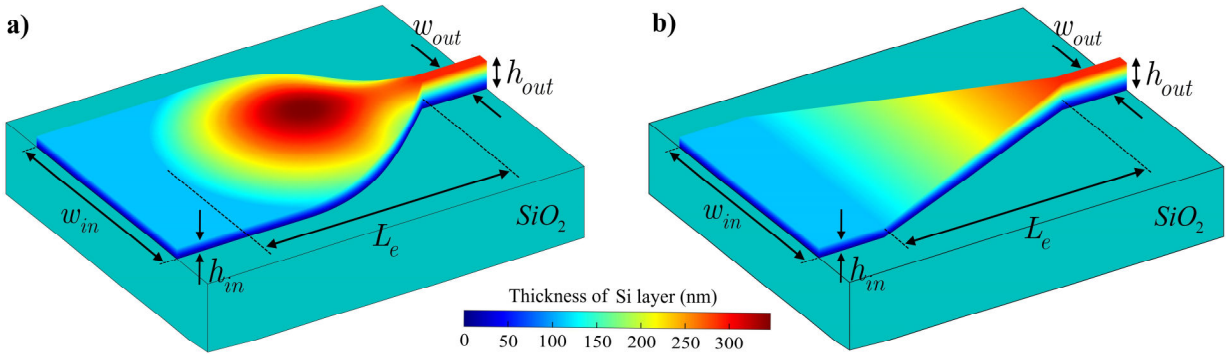

Fig. 11. The implementation of a) Taper E of Fig. 8(a) and b) linear taper with the same length of $L_e = 10.95\mu m$. These tapers are used to connect the input waveguide with $w_{in} = 10\mu m$ and $h_{in} = 110nm$ to the output waveguide with $w_{out} = 0.5\mu m$ and $h_{out} = 290nm$. The upper air cladding is not displayed.

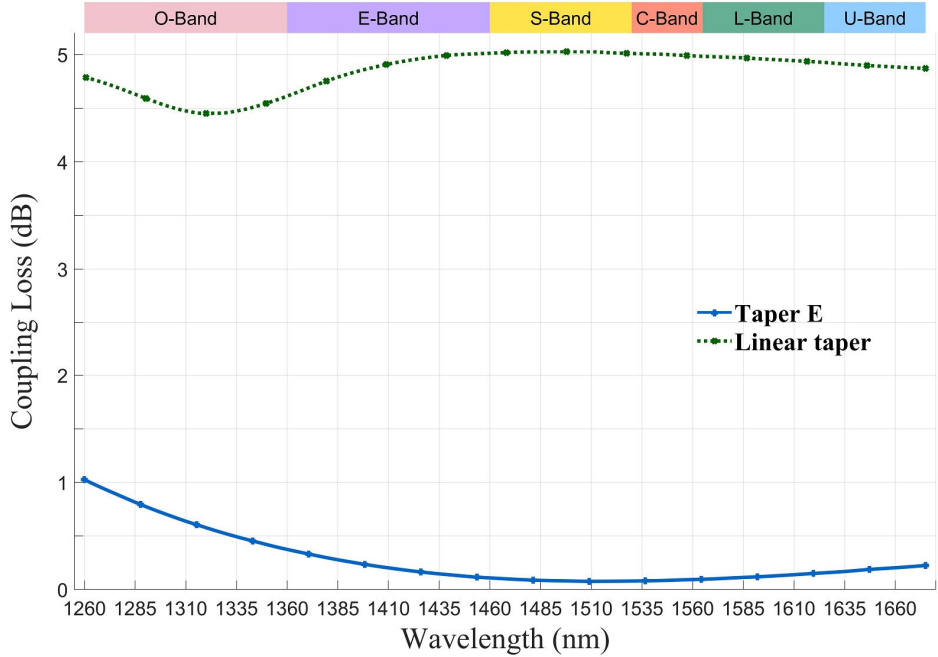

Fig. 12. Performance of the Taper E is compared with the linear taper of Fig. 11(b) for the TE$_0$ mode.

### 4.3 Comparison with previous studies

We compare Tapers A and E with previous studies in Table 1. Waveguide type, tapering mechanism, width of waveguides, length of the taper, coupling loss, bandwidth, and evaluation method are compared in this table. When available both theoretical and experimental results are presented. To distinguish between the theoretical and experimental results in the table, we present the experimental results in the parentheses. Our designed tapers have the shortest length compared to other studies. The designed tapers have a coupling loss of less than 1 dB in a broad bandwidth of 1260-1675 nm while the coupling loss is lower than 0.37 dB in the C-band which is comparable with the previous studies.

In general, some deviations are expected in the fabrication process of the designed tapers. The effects of such deviations on the performance of the designed tapers are numerically estimated. To study fabrication imperfection in the lens, the spatial Si thickness randomly ranging from -20 to +20 nm is added to the designed thickness of the ideal lens. The 3D numerical simulations indicate that introducing spatial Si thickness variation of +/- 20 nm to the truncated lens of Taper E results in the maximum excess loss of 0.3 dB in the C-band.

Table 1. Comparison of waveguide tapers

| Ref. (year) | Waveguide (WG) type | Tapering mechanism | initial-to-final width ($\mu m$) | Taper's Length ($\mu m$) | coupling loss (dB) | Bandwidth (nm) | Evaluation method |
|---|---|---|---|---|---|---|---|
| [15] (2014) | SOI | Linear | 12-to-0.5 | 120 | 0.07 | - | Theoretical |
| [16] (2017) | SOI | Flat-lens | 10-to-0.5 | 22.5 | <0.45 | 1520-1580 | Theoretical |
| [17] (2018) | grating WG-to-SiN WG | quadratic sinusoidal taper | 10-to-1.0 | 19.5 | <0.27 (<0.8) | 1480-1640 (1550-1630) | Theoretical (Experimental) |
| [18] (2017) | SOI | quadratic sinusoidal taper | 10-to-0.5 | 15.0 | <0.31 (>5.45) | 1500-1650 (1520-1580) | Theoretical (Experimental) |
| [19] (2014) | SOI | Genetic algorithm | 12-to-0.5 | 20.0 | <0.5 (<0.7) | 1520-1580 (1520-1565) | Theoretical (Experimental) |
| [20] (2018) | grating WG-to-SOI WG | Hollow tapered WG with Si strips | 15-to-0.3 | 60.0 | <2.0 | 1531-1578 | Theoretical |
| Taper A | SOI | Luneburg lens | 10-to-0.5 | 11.0 | <0.87 | 1260-1675 | Theoretical |
| Taper E | SOI | Luneburg lens | 10-to-0.5 | 10.95 | <1.02 | 1260-1675 | Theoretical |

## 5. Conclusion

Tapering between devices of varying dimensions is an area consuming function in PICs. Designing compact tapers with low coupling and return losses and broadband operation are essential in scaling down the size of PICs. We present the theoretical design of a compact taper with a length of 11 µm based on Luneburg lens which focuses the light from a 10 µm slab waveguide into a 0.5 µm single mode waveguide. Full-wave simulations verify that the coupling loss is lower than 0.87 dB in the entire O, E, S, C, L, and U bands of optical communications. Simultaneous vertical and lateral tapering is achieved by flattening the Luneburg lens through QCTO. A 10 µm × 0.11 µm waveguide is coupled to a 0.5 µm × 0.29 µm waveguide through a 10.95 µm-long taper with a coupling loss of less than 1.02 dB in the 1260-1675 nm bandwidth. The design procedure described in this paper can be easily expanded to couple waveguides with different geometries.